\documentclass[reprint,showpacs,preprintnumbers, amsmath,amssymb,aps,prc,]{revtex4-1}
\usepackage{amsfonts} 
\usepackage{amsmath}
\usepackage{amssymb}
\usepackage{graphicx,color}
\usepackage{rotating}
\usepackage{hyperref}
\usepackage{dcolumn}
\usepackage{bm,ulem}
\usepackage{subfig}
\usepackage{multirow}

\begin{document}

\title{Hyperons in the pasta phase}

\author{D\'{e}bora P. Menezes}
\affiliation{Depto de F\'{\i}sica - CFM - Universidade Federal de Santa
Catarina  Florian\'opolis - SC - CP. 476 - CEP 88.040 - 900 - Brazil}
\author{Constan\c ca Provid\^encia}
\affiliation{CFisUC, Department of Physics,
University of Coimbra, P3004 - 516  Coimbra, Portugal}
\email{cp@fis.uc.pt}

\begin{abstract}
We have investigated under which conditions hyperons (particularly
$\Lambda$s and $\Sigma^-$s) can be found in the pasta phase. The
larger the density and the temperature and the smaller the electron fraction the higher the probability that these particles
appear but always in very small amounts. $\Lambda$-hyperons only occur in the
gas and in smaller amounts than would occur if matter were homogeneous, never with abundancies above  10$^{-5}$.
 The amount of  $\Sigma^-$ in the
gas is at least two orders of magnitude smaller and can be disregarded in practical calculations.
\end{abstract}

\maketitle

\section{Introduction}

Not long ago, two massive stars were confirmed
\cite{Demorest,Antoniadis} giving rise to the hyperon puzzle: while
nuclear physics favours soft EOS at low densities,
massive stars can only be described by stiff EOS at high
densities. Meanwhile, some constraints were imposed to neutron stars
radii and it is now believed that the radii of canonical 1.4 solar
mass ($M_\odot$) pulsars are lower than what most of the models foresee
\cite{Lattimer_2016}. 

On the other hand, the observation of supernova explosions is not
trivial, a fact that contributed to the increasing importance of
simulations of core-colapse supernova and its remnants. An appropriate
EOS  for these simulations would range from very low to high densities
and from zero temperature to temperatures higher than 100 MeV, a not very easy
task to be accomplished. Hence, the very small number of these EOS in
the market, most of them publicly available in the CompOSE (CompStar
Online Supernovae Equations of State) database \cite{compose}.

Many attempts have been made to circumvent these two problems
and next we only mention some of the examples of the propositions
found in the literature. These problems can be tackled  
by choosing appropriate meson-hyperon couplings \cite{micaela_2016},
and introducing strange mesons as mediators of the baryons in
relativistic models \cite{Bednarek2012, Weissenborn2012,luiz1,luiz2} in already existing models or
by using combinations of different parts of models available in
the literature \cite{micaela_2017}. In all cases, the existence of a
degree of freedom that carries strangeness is important, see
\cite{vidana16} for a recent review.

In this context, the low density part of the EOS plays a role that
cannot be disregarded. The pasta phase results from the competition
between the strong and the Coulomb interactions at densities
compatible with the ones in the inner crust of neutron stars. Such
competition generates a frustrated system \cite{Avancini_2008,
  Xu_2009} and the name pasta
refers to specific shapes acquired by matter, namely: droplets,
bubbles, rods, tubes and slabs. The inclusion of the pasta phase in
the EOS practically does not
influence the stellar maximum masses, but certainly affects the radii \cite{Guilherme_2017,Lena_2016}.
So far, the pasta phase has only been
treated with nucleonic degrees of freedom.

From the considerations made above, it is obvious that the
constituents present in the equations of state (EOS) that describe 
neutron stars and supernova cores are the essential ingredients in the
determination of thermodynamic quantities and macroscopic properties.
Hence, the existence of the strangeness degree of freedom in the pasta
phase has to be investigated and this is the aim of the present
work. At zero temperature hyperons do not occur at densities below
two times the saturation density. However, at finite temperature the
appearance of hyperons is mainly governed by their mass and, the larger
the temperature the larger the probability that they appear at smaller
and smaller densities. We restrict our
work to finite temperature systems when the hyperons may occur at
densities typical of non-homogeneous matter.
We first consider the possibility that $\Lambda$ particles are present
in the pasta phase and check the conditions for their existence and
later discuss situations that could give rise to the onset of
$\Sigma^-$ as well. These are the strangeness carrying baryons that
usually appear first in stellar matter due to the values of their
masses. Typical density values of the hyperon onset in the 
homogeneous phase are given in the next section. The
 values encountered are used to justify the presence of hyperons in 
the pasta phase.

We next show only the most important formulae for the understanding of
our calculations and then display the results alongside some comments and 
conclusions.

\section{Formalism}

We have chosen to describe hadronic matter
within the framework of the relativistic non-linear Walecka model (NLWM)
\cite{walecka} with non-linear terms \cite{bb}.
 In this  model the nucleons are coupled to  the scalar $\sigma$, 
isoscalar-vector $\omega^\mu$ and isovector-vector $\vec \rho^\mu$  meson fields. 
We include a $\omega-\rho$ meson coupling term as in
\cite{nl3wr,fsu,iufsu,rafael_2011,Guilherme_2017} because this term was
shown to control the symmetry energy and its slope, resulting in
equations of state that can satisfy most of the nuclear matter
saturation properties and observational  constraints.
The Lagrangian density reads
\begin{eqnarray} 
{\cal L}&=& \sum\limits_{j{\kern 1pt}  = {\kern 1pt} 1}^4 {\bar \psi_j \left[ ~ { \gamma_\mu  \left( {i \partial^\mu 
- g_{\omega j} \, \omega^\mu - g_{\rho j} \, \vec \tau _j \,.\, \vec \rho^{\, \mu}  } \right) 
- m_j^* ~} \right]\psi_j }  \nonumber \\ 
&+& \frac{1}{2} {\partial_\mu \sigma \partial^\mu \sigma - \frac{1}{2} m_{\sigma}^2 \sigma^2 }  
- \frac{1}{{3!}} k \sigma^3  - \frac{1}{{4!}} \lambda \sigma^4     \nonumber \\
&-& \frac{1}{4} \Omega_{\mu \nu} \, \Omega^{\mu \nu}
+ \frac{1}{2} m_{\omega}^2 \, \omega_\mu \omega^\mu 
\nonumber \\
&-& \frac{1}{4} \vec R_{\mu \nu } \,.\,  \vec R^{\mu \nu } + 
\frac{1}{2} m_\rho^2 \, \vec \rho_{\mu} \,.\, \vec \rho^{\, \mu}   \nonumber \\
&+& \Lambda_{\rm v} (g_{\rho}^2 \; \vec \rho_{\mu} \,.\, \vec \rho^{\, \mu} )(g_{\omega}^2 \; \omega_{\mu} \omega^{\mu} )  \;,   
\label{baryon-lag}   
\end{eqnarray}
where $m_j^* = m_j - g_{\sigma j}\, \sigma$ is the baryon effective mass,  
$\Omega_{\mu\nu }=\partial_\mu \omega_\nu - \partial_\nu \omega_\mu$~, 
$\vec R_{\mu \nu } = \partial_\mu \vec \rho_\nu -\partial_\nu \vec \rho_\mu 
-g_\rho \left({\vec \rho_\mu \,\times \, \vec \rho_\nu } \right)$, $g_{ij}$ 
are the coupling constants of mesons $i = \sigma, \omega, \rho$ with baryon 
$j$, $m_i$ is the mass of meson $i$. 
The couplings 
$k$~($k = 2\,M_N\,g_{\sigma}^3\,b$) and $\lambda$~($\lambda = 6\, g_{\sigma}^4\,c$) 
are the weights of the non-linear scalar terms and $\vec \tau$ is the isospin operator. 
The sum over $j$ extends over the  lightest 4 baryons 
($n,p,\Lambda,\Sigma^-$). In the present work we have opted to use
the NL3$\omega\rho$ parametrisation \cite{nl3wr}, which is an extension of the NL3
parametrisation \cite{nl3} with the inclusion of the $\omega-\rho$
interaction.  For the hyperon-$\omega$ interaction we take SU(6)
symmetry, and for the hyperon-$\sigma$ interaction we consider that
the $\Lambda$, $\Sigma$ and $\Xi$ potentials in symmetric nuclear
matter at saturation are, respectively, -28 MeV, +30 MeV and -18 MeV. 
The parameters are: $m_\sigma = 508.194$ MeV,
$m_\omega=782.501$ MeV, $m_\rho=763$ MeV, $g_{\sigma n}=10.217$ ($n$
stands for protons and neutrons), $g_{\sigma \Lambda}=6.323$, 
$g_{\sigma \Sigma}=4.708$, $g_{\omega n}=12.868$, $g_{\omega
  \Lambda}=g_{\omega \Sigma} =8.578$, $g_{\rho j}=11.276$, 
$k/M= 2 \times 10.431$,
$\lambda=-6 \times 28.885$ and the corresponding saturation properties are:
density at  $0.148$ fm$^{-3}$, binding energy of -16.2 MeV,
  compressibility equal to 271.6 MeV, symmetry energy  of 31.7 MeV and 
  slope equal to 55.5 MeV.  From the results in \cite{fortin16}, one
  can see that this model predicts stellar masses above 2$M_\odot$ and 
satisfies several presently accepted  experimental and theoretical constraints.

The pasta phase is obtained for charge neutral matter and leptons are
usually incorporated because their presence is expected both in the
interior of neutron stars and in the core-colapse supernova. The 
leptonic Lagrangian density is simply

\begin{equation} 
{\cal L}= {\bar \psi _l \left( {i\gamma _\mu  \partial ^\mu 
- m_l } \right)\psi _l},
\end{equation}
where $l$ represents only the $e^-$ in the present work, whose mass is
0.511 MeV. The leptons enter the calculations only via the weak interaction.

The construction of the pasta phase obeys the well known Gibbs
conditions for phase coexistence and in the present work we opt for
the coexistence phase (CP) method extensively discussed in previous works 
\cite{Avancini_2008,thomasfermi,alfas,clusters, Guilherme_2017}, which we do not
repeat here. The particle chemical potentials are defined in terms of
a baryon chemical potential ($\mu_B$) and a charge chemical potential ($\mu_Q$),
which are the quantities enforced as identical in both phases, such as
\begin{equation}
\mu_j=\mu_B + q_j \mu_Q,
\end{equation}
where $q_j$ is the electric charge of each particle.
The electron fraction is fixed by the imposition of charge neutrality:
\begin{equation}
Y_e=\frac{\rho_e}{\rho} = Y_Q=\frac{\rho_Q}{\rho} , \quad \rho_Q=\rho_p-\rho_\Sigma
\end{equation}

We also define the fraction of $\Lambda$ and $\Sigma^-$ particles as:
\begin{equation}
Y_{\Lambda_1}=\frac{\rho_{\Lambda_1}}{\rho} ,\quad
Y_{\Lambda_2}=\frac{\rho_{\Lambda_2}}{\rho} ,
\label{fraclglobal}
\end{equation}
\begin{equation}
Y_{\Sigma_1}=\frac{\rho_{\Lambda_1}}{\rho} ,\quad
Y_{\Sigma_2}=\frac{\rho_{\Lambda_2}}{\rho} ,
\label{fracsiglobal}
\end{equation}
where the subscript $1$ refers to the dense phase and $2$ to the gas
phase and $\rho=\rho_p+\rho_n+\rho_\Lambda+\rho_\Sigma$. In most
cases studied in this work the $\Sigma^-$ particles are not
present. In this case, the density is given only by
$\rho=\rho_p+\rho_n+\rho_\Lambda$ and $\rho_Q=\rho_p$.

An important aspect generally discussed is the surface tension
coefficient ($\sigma$). We use the $\sigma$ parametrisation given in \cite{clusters}.

 Before we discuss the presence of hyperons in the pasta phase, it
  is important to investigate their onset in homogeneous matter.
We start by analysing the two usually considered scenarios at
  zero temperature. The first one refers to stellar matter, where the
  equation of state is
obtained with the assumption of charge neutrality and
$\beta$-equilibrium and the fraction of leptons includes electrons and
muons, varies with density and it is an
  output of the calculation. In the second scenario, we have
  considered the electron
 fraction as a fixed quantity that enters as an input, as in the finite temperature
 case examined throught out this work. In this case, no muons are
 incorporated in our calculations.  As seen in table \ref{tab1},
 no hyperons appear at sub-saturation density at $T=0$, as already
 expected. 
It is interesting to notice the competition between the
 contribution from the $\rho$-meson that in asymmetric matter favors
 the hyperons with the  smallest charges, and the $\sigma$-meson that
 favor the hyperons with an attractive potential in symmetric nuclear
 matter at saturation.  In symmetric matter, the $\rho$-meson
 contribution is zero, and the hyperons of each isospin multiplet with
smaller mass are favored.
 
 In table \ref{tab2} we display the onset of hyperons at
  different temperatures  and electron fractions.  We have considered
  the three hyperons with a larger fraction at low densities and we
  are interested in the occurence of hyperons at densities below 0.1
  fm$^{-3}$.  From the table we see that this is only possible for
  $T>5$ MeV, taking as reference a hyperon fraction larger than
  $10^{-12}$. While at {very} low $T$  the sequence of hyperons may be
  different, at { higher} $T$ the sequence is generally $\Lambda$,
  $\Sigma$ and $\Xi$. For a large electron fraction, close to $Y_e=0.5$
  the hyperon of the $\Sigma$  triplet or $\Xi$ doublet  with the
  largest charge is the first to appear due to the smaller mass and
  the small contribution of the
  $\rho$-meson. In fact, at $Y_e=0.5$ the members of each isospin
  multiplet come very close together, being only distinguished by the
  mass. A small value of $Y_e$ originates a large contribution from the
  $\rho$-meson and the hyperon with the most negative isospin
  projection is the favoured one, i.e.  $\Sigma^-$ and $\Xi^-$. 
  Temperature washes quite fastly
  the differences due to the different optical interactions taken at
  $T=0$ and therefore, the $\Sigma$ hyperon with a smaller mass than
  the $\Xi$  hyperon ends up being favored.  The
  $\Lambda$-meson is always the one with the largest abundancy, and we
  next concentrate our study in the occurence of this hyperon.  For
  reference, we also show some results for $\Sigma$ but from table \ref{tab2}
  we conclude that when they occur at densities of the pasta phases
  they are 2 to 3 orders of magnitude less abundant than the
  $\Lambda$s. The $\Xi$-hyperon fractions are negligible for the
  densities and temperatures where pasta phases occur.

\begin{table}[ht!]
\begin{center}
{\footnotesize
 \begin{tabular}{ccccccc}
   \hline 
$Y_e$ & $\rho_{\Lambda}$&
   $\rho_{\Sigma^+}$ &  $\rho_{\Sigma^0}$ &  $\rho_{\Sigma^-}$ &
  $\rho_{\Xi^0}$ & $\rho_{\Xi^-}$  \\ 
(MeV) & (fm$^{-3}$) & (fm$^{-3}$) & (fm$^{-3}$) & (fm$^{-3}$) & (fm$^{-3}$) & (fm$^{-3}$)\\
   \hline 
 $\beta$-eq & 0.31 & - &- & - & 0.64 & 0.35 \\
  0.1 & 0.30 & - & - & - & 0.61 & 0.34 \\
  0.30 & 0.33 & 0.77 & - & - & 0.60 & - \\
  0.50 & 0.36 & 0.60 & - & - & 0.57 & - \\
 \hline
\end{tabular}}
\end{center}
\caption{ Hyperon onset densities obtained at zero 
   temperature.}
\label{tab1}
\end{table}

\begin{table}[ht!]
\begin{center}
{\small
\begin{tabular}{c cc cc cc cc}
\hline
$T$& $\rho_\Lambda$& $Y_\Lambda$& $\rho_{\Sigma^-}$& $Y_{\Sigma^-}$&
                                                                     $\rho_{\Xi^-}$& $Y_{\Xi^-}$\\
(MeV) & (fm$^{-3}$) &  & (fm$^{-3}$) &  & (fm$^{-3}$) & \\
\hline
$Y_e=0.1$&$\Lambda$& & $\Sigma^-$& & $\Xi^-$&\\
0.001& 0.31   &    $10^{-12}$ & & -    $<10^{-12}$ &  0.34   &    $10^{-12}$ \\
1&    0.28    &    $10^{-12}$  &   0.33    &     $10^{-12}$    & 0.32&  $10^{-12}$ \\
3 &   0.23    &    $10^{-12}$ &    0.28    &    $10^{-12}$  & 0.29    &    $10^{-12}$   \\
5 &   0.13    &    $10^{-12}$ &   0.24   &    $10^{-12}$   & 0.26     &    $10^{-12}$  \\
7&    0       &    $10^{-11}$  &     0.18   &    $10^{-12}$   & 0.22     &    $10^{-12}$   \\ 
9 &   0      &    $3.\times 10^{-9}$   &     0     &    $5.\times 10^{-12}$        & 0.17    &    $10^{-12}$      \\ 
10&   0       &    $2. \times 10^{-8}$  &    0           &    $9.\times 10^{-11}$      & 0.14    &    $10^{-12}$ \\
12&    0      &    $5.\times 10^{-7}$   &   0    & $6.\times 10^{-9}$  & 0.03 &$10^{-12}$ \\
14 &    0      &    $4.\times 10^{-7}$   &   0    & $1.\times 10^{-7}$  & 0 &$2.\times 10^{-11}$ \\
\hline
\\
$Y_e=0.3$ &$\Lambda$& & $\Sigma^-$& & $\Xi^-$&\\
0.001 &  0.33&    $10^{-12}$   & -   &     $<10^{-12}$      & - &  $<10^{-12}$\\  
1    & 0.31 &    $10^{-12}$   & -     &     $<10^{-12}$       & 0.56 &    $10^{-12}$ \\
3   & 0.26  &    $10^{-12}$    & -    &     $<10^{-12}$        & 0.40 &    $10^{-12}$ \\  
5    & 0.24 &    $10^{-12}$       & 0.34 &    $10^{-12}$  & 0.33 &    $10^{-12}$\\
7   &  0   &$10^{-11}$            & 0.29  &    $10^{-12}$ & 0.29  &    $10^{-12}$\\ 
9   &  0  &$3.\times 10^{-9}$             & 0.21  &    $10^{-12}$  & 0.26 &    $10^{-12}$ \\ 
10  & 0  &$2.\times 10^{-8}$         &    0    &$2\times 10^{-11}$        & 0.23 &    $10^{-12}$\\
12  & 0   &$4\times 10^{-7}$         &   0   &$10^{-9}$         &0.16  &    $10^{-12}$\\ 
14  & 0   &$3\times 10^{-6}$         &   0   &$3.\times 10^{-8}$
                                                                   &0
                                                                                   &    $4.\times 10^{-12}$\\ 
\hline
\\
$Y_e=0.5$ &$\Lambda$& & $\Sigma^+$& & $\Xi^0$&\\
.001      & 0.37 &    $10^{-12}$ & 0.62  &     $10^{-12}$         & 0.60 &    $10^{-12}$ \\
1           & 0.35 &    $10^{-12}$& 0.59  &     $10^{-12}$         & 0.57  &    $10^{-12}$ \\
3            & 0.30 &    $10^{-12}$  &0.53  &     $10^{-12}$         & 0.50 &    $10^{-12}$ \\
5            & 0.24  &    $10^{-12}$ & 0.47   &     $10^{-12}$        & 0.43 &    $10^{-12}$    \\
7              & 0  &    $8.\times 10^{-12}$      & 0.37 &     $10^{-12}$         & 0.38  &    $10^{-12}$  \\
9              & 0   &    $2.\times 10^{-9}$      & 0.34 &    $10^{-12}$    & 0.34  &    $10^{-12}$ \\
10           & 0    &    $ 10^{-8}$      & 0.31   & $ 10^{-12}$    & 0.32  & $10^{-12}$\\
12           & 0    &    $2.\times 10^{-7}$       & 0  &   $ 6.\times 10^{-10}$       & 0.27 &    $10^{-12}$ \\
14           & 0    &    $2.\times 10^{-6}$       & 0  &   $ 3.\times
                                                         10^{-9}$
                                                                   & 0
                                                                                   &    $10^{-12}$ \\
\hline
\end{tabular}
}

\end{center}
\caption{ 
Hyperon onset densities for hyperon fraction equal or above
$10^{12}$. Only results for $\Lambda$, $\Sigma^-$ and $\Xi^-$ are
shown. For $Y_e=0.5$ the  $\Xi^0$, $\Sigma^+$ and
$\Sigma^0$ appear before, respectively, $\Xi^-$ and  $\Sigma^-$   but
the differences are small and the fractions are always very  small. }
\label{tab2}
\end{table}

\section{Results and discussions}

\begin{figure}[ht]
\centering
\includegraphics[width=.9\linewidth]{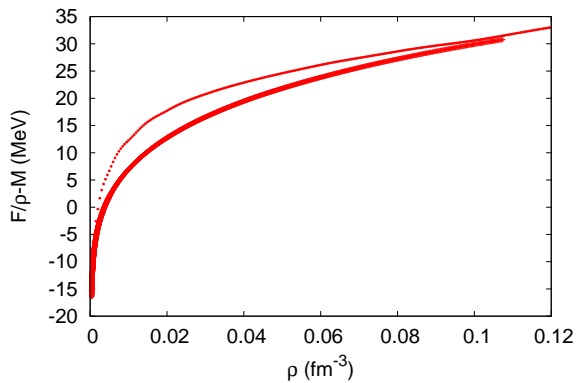}
\caption{Free energy versus baryon density for homogeneous matter
  (dotted line) and for the pasta phase (solid line) for $T=10$ MeV
  and $Y_e=0.3$} 
\label{fig1}
\end{figure}

In the present section, we discuss under which conditions the fraction
of hyperons, in particular, of $\Lambda$s, is largest in the range of
densities where pasta phases occur. The calculations are performed
within the formalism presented in the last section. Although in the CP
approach to the pasta phases, the surface
energy and the Coulomb energy are added after the minimisation of the
free energy, we consider it is enough to get the correct idea of the
amount of hyperons that occur in the non-homogeneous
sub-saturation warm stellar matter.  We perform the study within the
NL3$\omega\rho$ parametrisation described in the last section.

\begin{figure}
\centering
\begin{tabular}{c}
\subfloat[]{\includegraphics[width=.8\linewidth]{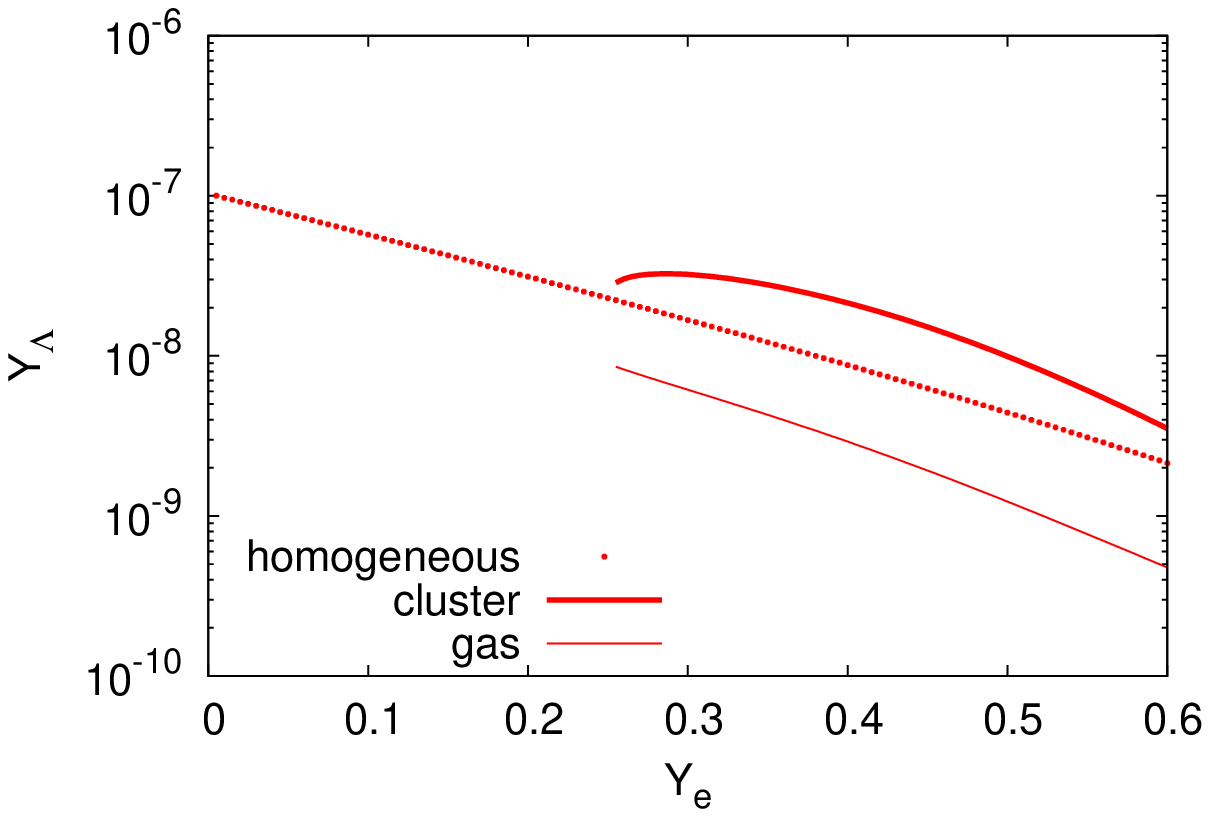}\label{fig2:a}}\\
\subfloat[]{\includegraphics[width=.8\linewidth]{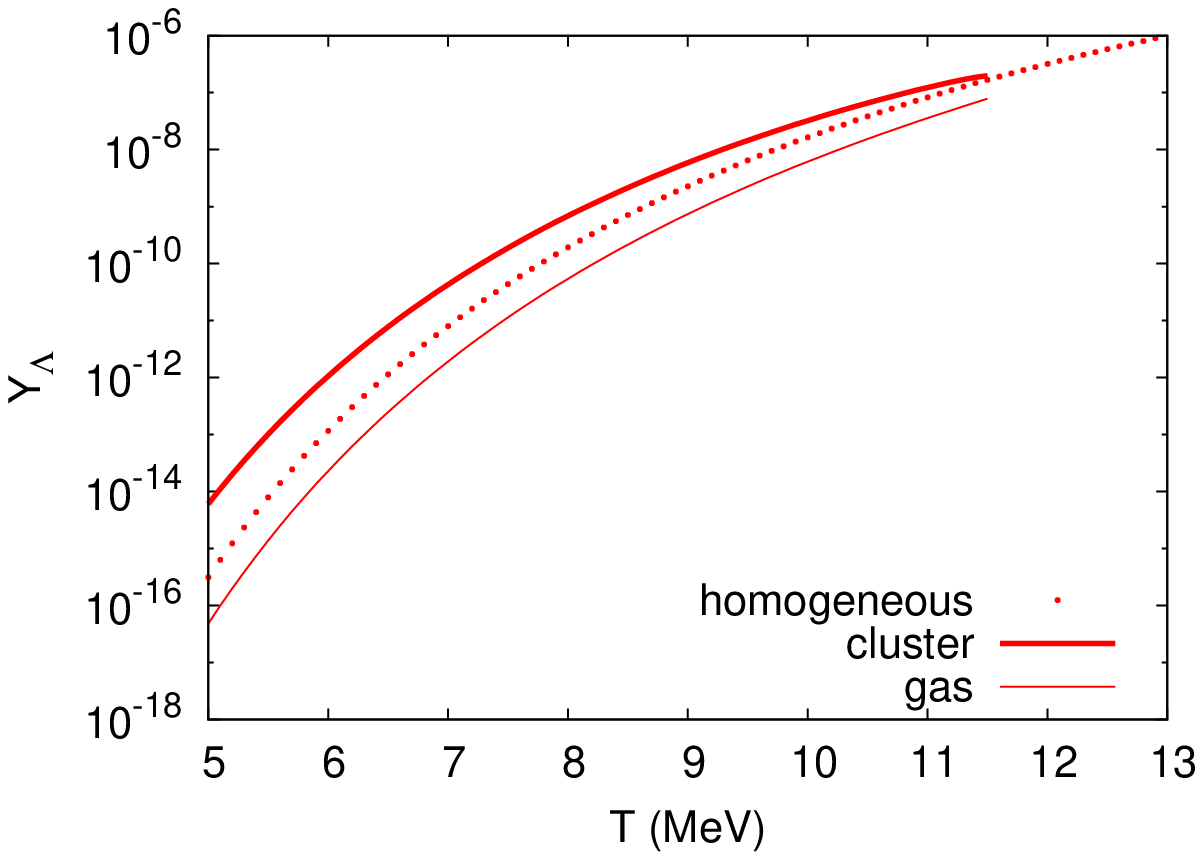}\label{fig2:b}}\\
\end{tabular}
\caption{$\Lambda$ fractions obtained with  with $\rho=0.05$ fm$^{-3}$
  as function of (a) the charge fraction at T=10 MeV and (b) the
  temperature with $Y_e=0.3$.}
 \label{fig2}
\end{figure}

We illustrate how the free energy per particle decreases when
non-homogeneous matter is considered instead of homogeneous matter in
Fig.\ref{fig1} taking $T=10$ MeV and $Y_e=0.3$. The range of densities where the non-homogeneous
matter occurs varies with temperature and electron fraction. In
particular, it decreases as the temperature increases and eventually
disappears above a certain critical temperature, which is around 14.45
MeV for this model \cite{alam17}. Moreover, the electron fraction has a strong effect in the
extension of the pasta phase: since stellar matter is neutral, the larger the electron
fraction the larger the proton fraction and the larger the
non-homogeneous matter extension. 

\begin{figure}[ht]
\centering
\begin{tabular}{c}
\subfloat[]{\includegraphics[width=0.8\linewidth]{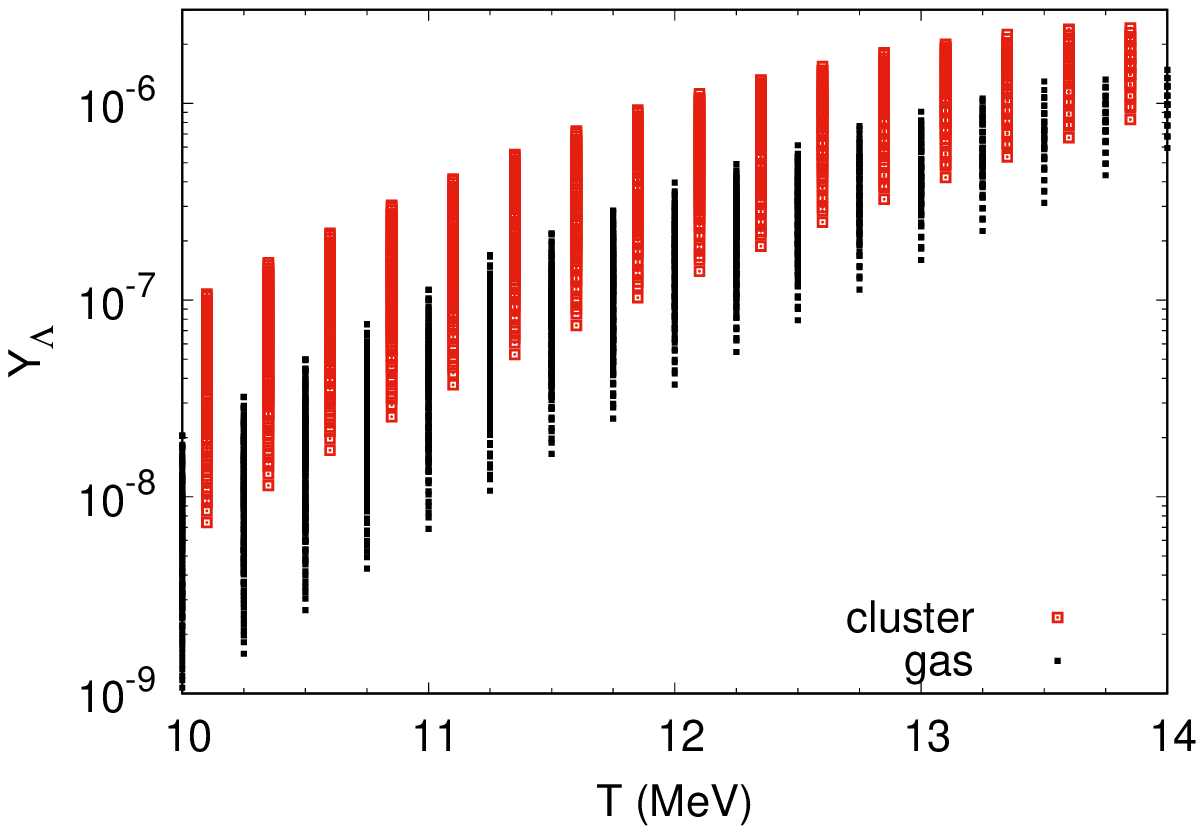}}\\
\subfloat[]{\includegraphics[width=0.8\linewidth]{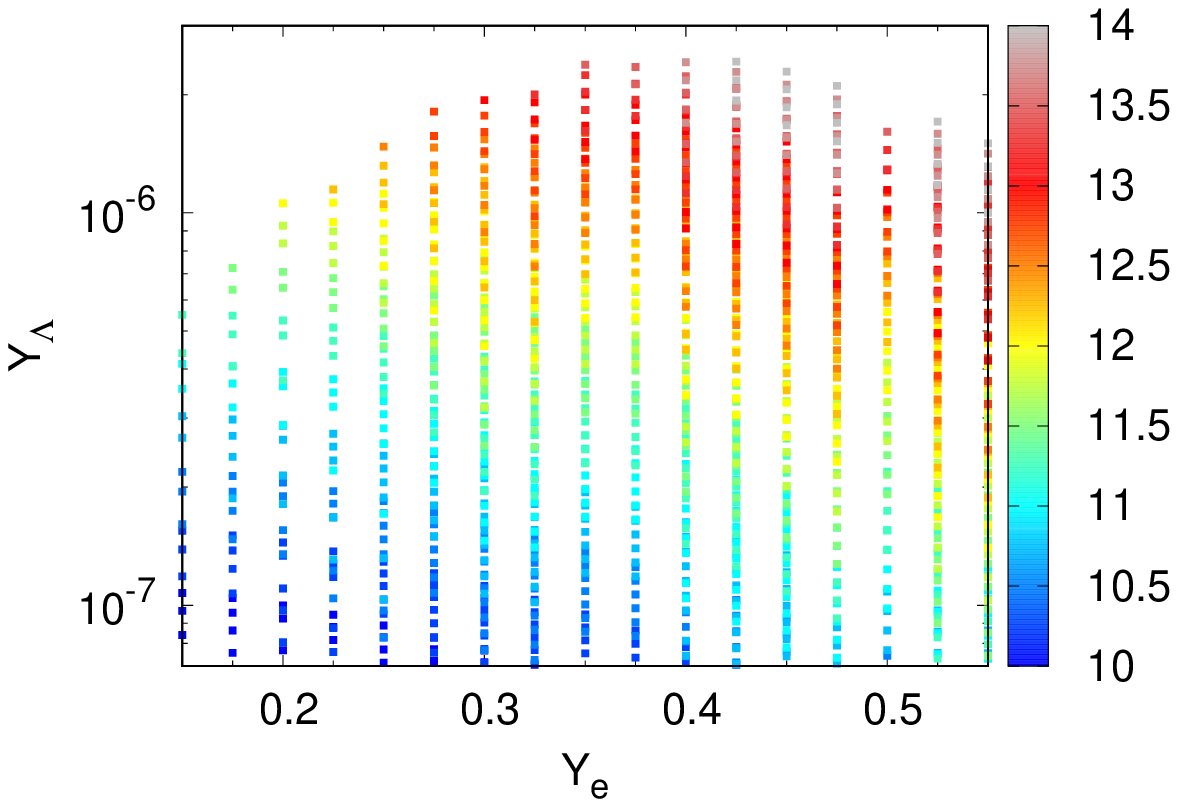}}\\
\subfloat[]{\includegraphics[width=0.8\linewidth]{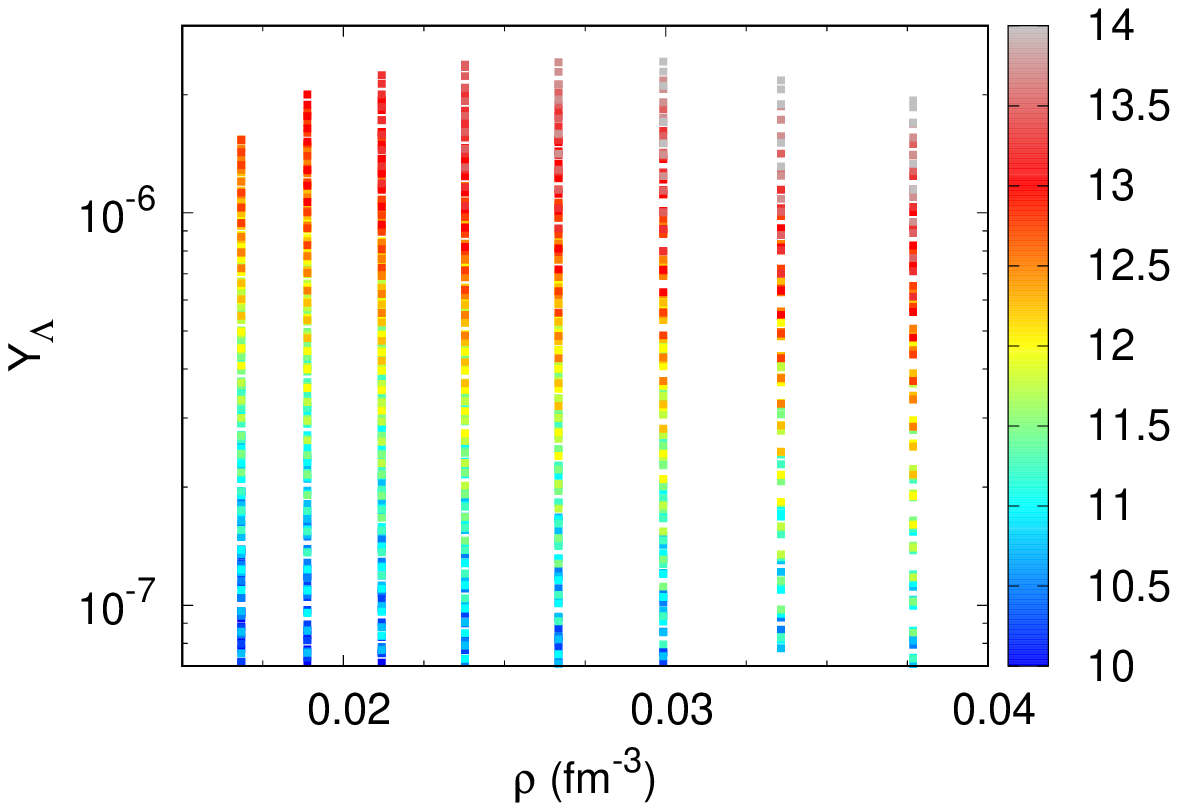}}\\
\end{tabular}
\caption{$\Lambda$ fraction as a function of (a) temperature in the
  clustered (red) and gas (black) phases; (b)
  electron fraction; (c) density under the conditions that predict
  fractions above $10^{-7}$. In (a),  to improve the visibility,  the
  temperature of the clusters  was shifted to the right by $\Delta T = 0.1$ MeV. In (b) and (c) only the $\Lambda$
  fractions in the clustered phase are shown, and temperature (in MeV)
  is
  indicated by a color index.} 
\label{fig3}
\end{figure}

In Fig. \ref{fig2} the fraction of $\Lambda$s is plotted as a function
of  (a)  the electron fraction $Y_e$  for $T=10$ MeV and (b) the
temperature for $Y_e=0.3$. The
dotted lines represent the fraction of $\Lambda$s that would occur  in
homogeneous matter. In (a) the thick and the thin lines are the fractions of
$\Lambda$s with respect to the total density in the dense (cluster) and gas
phases, as given in eq.(\ref{fraclglobal}). Since there is no distillation effect for strangeness as there is
for isospin \cite{gulminelli13}, the larger/smaller fraction in the
dense/gas phases simply reflects the fact that the larger the density
the larger the probability that hyperons occur. Notice that under the
conditions for pasta to occur these fractions are really very small.
We may also conclude that the clusters contain no hyperons since the
fraction is so small that it is not enough to predict a whole hyperon
inside the clusters. This implies that in the non-homogenous phase
hyperons only occur in the gas and in smaller amounts than would occur
if matter were homogeneous. This is illustrated in Fig \ref{fig3}(a)
where the black marks refer to the gas phase and the red ones to the
dense phase.

\begin{figure}[ht]
\centering
\begin{tabular}{c}
\includegraphics[width=0.9\linewidth]{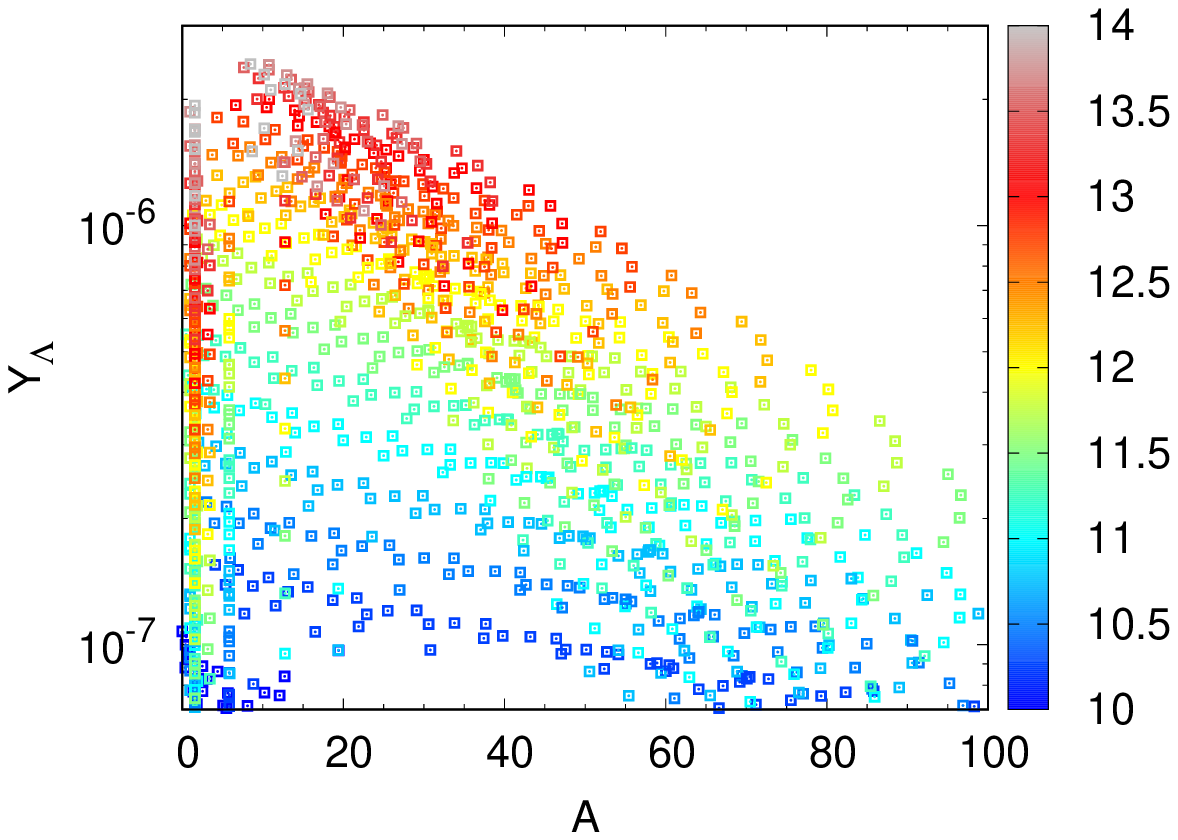}\\
\end{tabular}
\caption{$\Lambda$ fraction as a function of the number of nucleons in
  the heavy clusters under the conditions that predict
  fractions above $10^{-7}$. The temperature (in MeV)
  is
  indicated by a color index.} 
\label{fig4}
\end{figure}

Since the amount of strangeness seems to be so small we have
determined which are the  conditions that most favor the appearance of
hyperons, taking the temperature between 4 and 14 MeV, the electron
fraction between 0.05 and 0.6 and sub-saturation densities, see Fig. \ref{fig3}. Hyperon
fractions above $10^{-7}$ were possible only for $T>10$ MeV.  At these
temperatures, pasta phases do not occur for too high or too low
densities, neither for too small electron fractions.  However, the
smaller the electron fraction the larger the $\Lambda$ fraction at a
given temperature, which is distinguishable by a color index, since
 these conditions favour the replacement of  neutrons by  $\Lambda$s
 and decrease the free energy density.

There are, in fact, two competing factors related
  to the electron fraction: while more hyperons are favored with a smaller electron
  fraction to release the neutron pressure, the pasta extension is
  smaller for a smaller value of $Y_e$. As a result, it is not
  possible to attain so large temperatures with a small value of
  $Y_e$, and this explains the decrease of the $\Lambda$ fraction with
  a decrease of $Y_e$ if $Y_e<0.35$ as seen in
  Fig \ref{fig3}b).

 The internal structure of
   the pasta phase depends on the density, temperature and amount of
   charged particles, and for $Y_s > 10^{-7}$ practically only
   droplets survive.

\begin{figure}[ht]
\begin{center}
\begin{tabular}{c}
\subfloat[]{\includegraphics[width=0.75\linewidth]{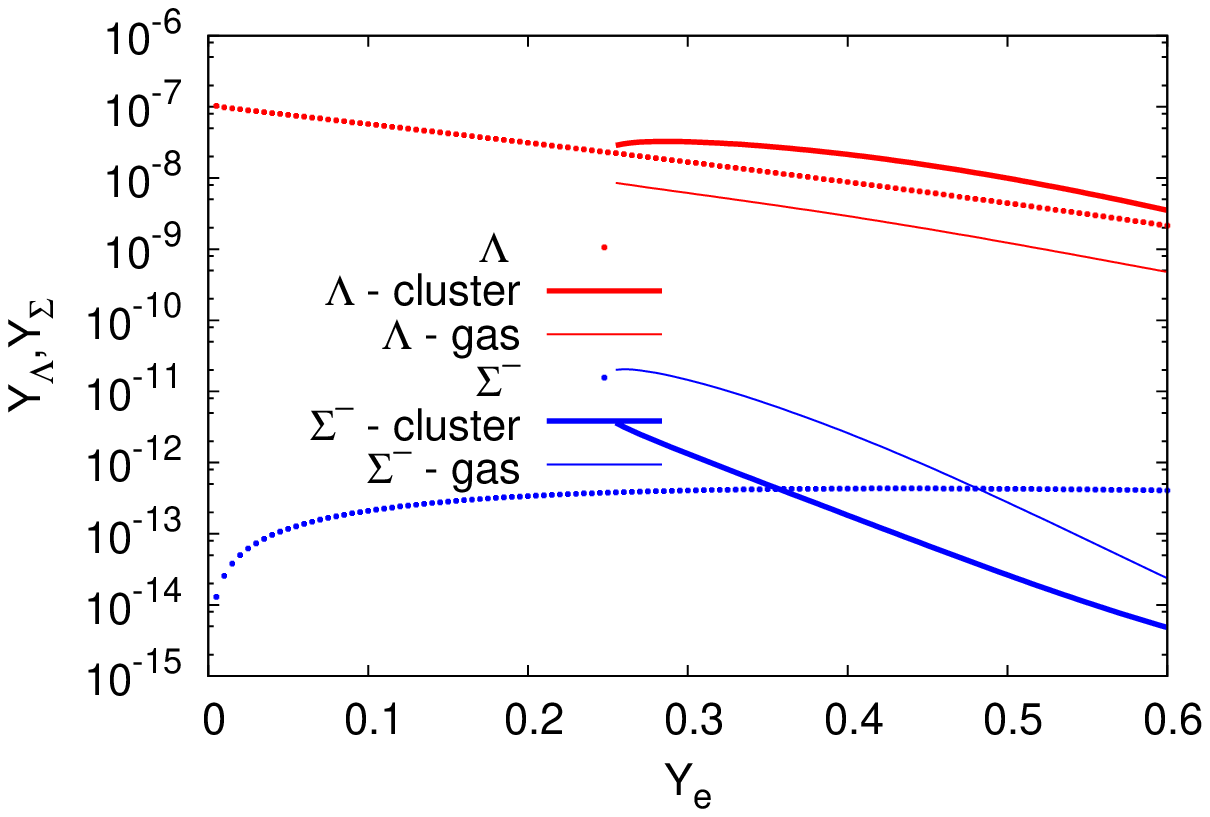}\label{fig5:a}}\\
\subfloat[]{\includegraphics[width=0.75\linewidth]{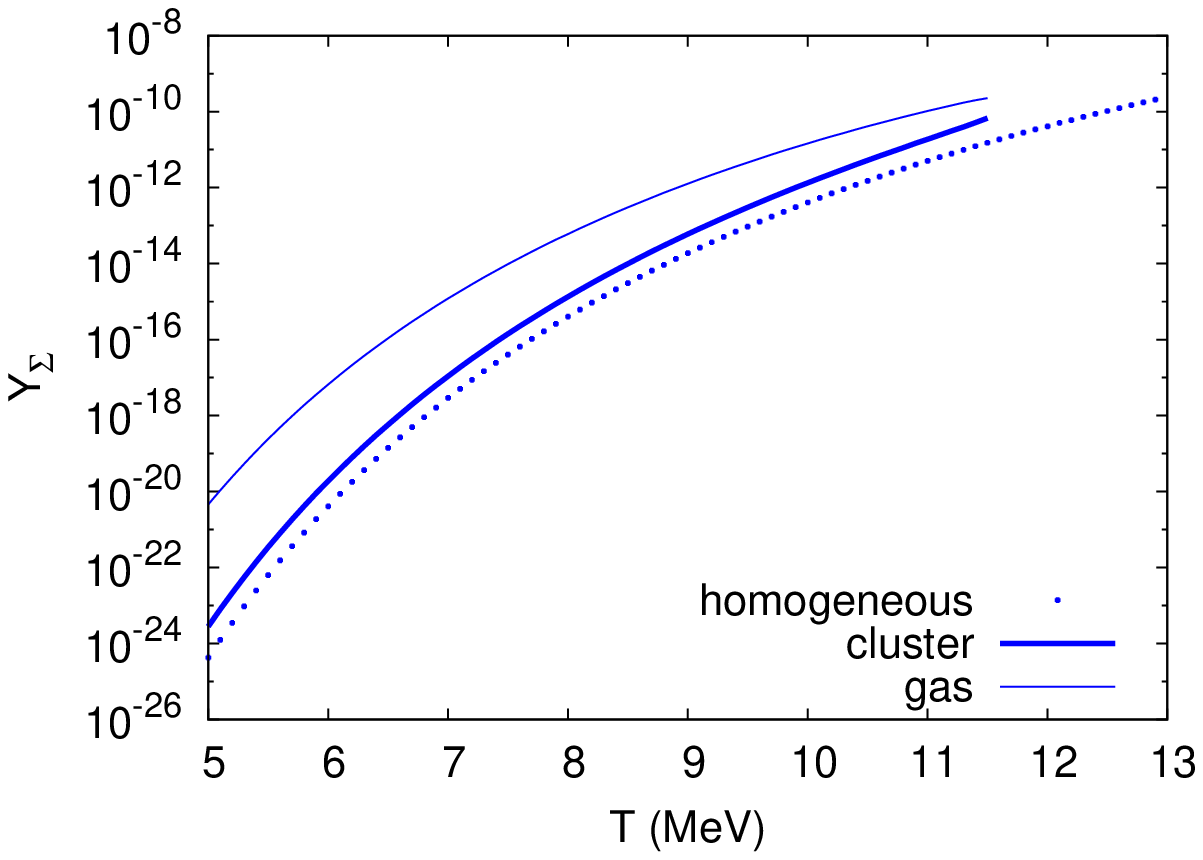}\label{fig5:b}}\\
\subfloat[]{\includegraphics[width=0.75\linewidth]{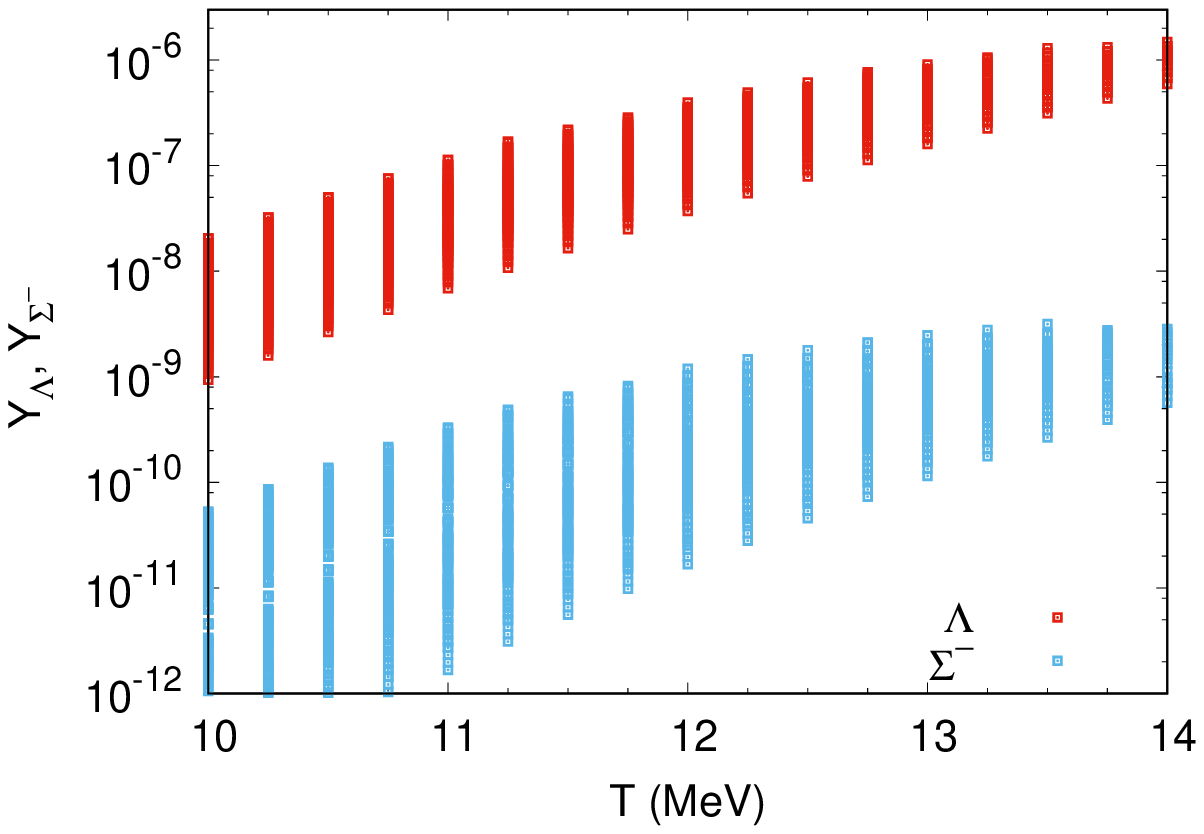}\label{fig5:c}}\\
\end{tabular}
\end{center}
\caption{Hyperon fractions obtained with $\rho=0.05$ fm$^{-3}$
  as function of a) the charge fraction at T=10 MeV and b) $Y_\Sigma$
  obtained as a function of the   temperature with $Y_e=0.3$.  In  c)
  the fractions of $\Lambda$s and $\Sigma^-$s are given as a function
  of temperature for a range of densities between 0.015 and 0.035
  fm$^{-3}$ and electron fraction $Y_e$ between 0.15 and 0.5.
Blue lines refer to $\Sigma$s  and red lines to $\Lambda$s.}
\label{fig5}
\end{figure}

In the present calculation only heavy clusters have been taken into account.
The size of the clusters that occur under the conditions that favour
hyperons can give an important information. In
 Fig. \ref{fig4}, we illustrate the number of nucleons in the
clusters for the $\Lambda$ fractions above $10^{-7}$. The largest
fractions occur precisely for the smaller clusters. On the other hand
the present approach is not good enough to describe the
non-homogeneous matter at the boundary to the core and at the low
density where the light clusters are most probable. The figure
indicates that a study similar to the present one should be performed
considering explicitly light clusters.  Under these conditions larger
fractions of $\Lambda$s as obtained in \cite{oertel17} are expected.

For the sake of completeness, we also allow for the presence of
$\Sigma^-$ particles and their fraction is plotted in Fig.\ref{fig5}
as a function of  (a) the electron fraction  for $T=10$ MeV alongside the fraction
of $\Lambda$s, (b) the temperature for $Y_e=0.3$ and (c) the
temperature for a range of densities between 0.015 and 0.035
  fm$^{-3}$ and electron fraction $Y_e$ between 0.15 and 0.5.
 In Figs. \ref{fig5}(a) and \ref{fig5}(c), the
$\Lambda$ fraction is also displayed, so that the individual fraction
of hyperons can be more easily compared.
As seen from this figure and expected from the previous
discussions, the amount of $\Sigma^-$s is almost negligible and
can be disregarded in practical calculations. The $\Sigma^-$ hyperons
are the second to occur due to their charge and mass. Even
though the repulsive potential of $\Sigma$s  in nuclear matter
disfavours their appearance at finite temperature, for the small
densities we consider the interaction plays a secondary role.
The different behaviour of the fraction of $\Lambda$s and $\Sigma^-$s
  in non-homogeneous and homogeneous matter
  can be attributed to the different  isospin character of these hyperons: a) the
  $\Lambda$s are not sensitive to isospin and their abundance is
  determined by the density, therefore a larger fraction is expected
  in denser matter;  b) $\Sigma^-$ has isospin projection -1 and is favoured in
  asymmetric nuclear matter as the one occurring in the background gas
  of the non-homogeneous matter. This explains why the fraction of
  $\Sigma^-$s is larger in the gas phase. In Fig. \ref{fig5}(c) we
  compare the abundances of  $\Lambda$s and $\Sigma^-$s in the
  background gas, in the conditions that most favour the appearance of
  these hyperons in the core-collapse supernova matter.  It is seen
  that the $\Sigma^-$s  fraction are essentially two orders of
  magnitude smaller than the  $\Lambda$ ones.

\section{Conclusions}

We report a study on the presence of hyperons in the
non-homogeneous phase of core-collapse
supernova matter. This was performed within the framework of a RMF EoS
with properties compatible with the ones presently accepted. 
The non-homogeneous phase was described within a coexisting phase
approach which does not take into account in a self-consistent way the
finite size effects. However, the results obtained within this
approach above $\rho\sim 0.01$ fm$^{-3}$ give a prediction not far
from a self-consistent Thomas Fermi calculation \cite{shen14}, and it
is within  this range of densities that the hyperons most contribute.

We have shown that the contribution of hyperons to the non-homogeneous 
matter is generally negligible: 
the fraction of $\Lambda$s obtained in all ranges of temperatures,
densities and electron fraction is always below $10^{-5}$. The
largest fractions occur for temperatures above 10 MeV, electron
fractions between 0.3 and 0.5 and densities between 0.025 and 0.035 fm$^{-3}$.
 Other hyperons such as the $\Sigma^-$ occur with two to six orders of
magnitude smaller. 

One interesting conclusion is that the heavy clusters  carry no
hyperons and the fraction of $\Lambda$s in the gas phase is smaller
than the expected fraction in homogeneous matter at the same
density because these hyperons are not sensitive to the isospin
distillation effect. On the other hand $\Sigma^-$s are sensitive to
isospin but in the best conditions their abundance is two to three
orders of magnitude below the $\Lambda$s abundance.
This seems to indicate that the role of hyperons in
matter with heavy clusters is negligible and can be taken into account
by properly including hyperons in the background gas. A different problem
concerns the appearance of hyperons together with light clusters which
in the present study were not included: it was shown that the largest
amounts of hyperons occur precisely with the smaller heavy
clusters. Since our approach fails in the region of light clusters a
calculation taking these degrees of freedom into account should be performed.

\section*{Acknowledgements}
C.P. thanks the warm hospitality at Universidade Federal de Santa
Catarina where this work was started. C.P. would also like to
acknowledge discussions on the appearance of hyperons at low densities
with F. Gulminelli and M. Oertel.
This work was partially supported by Conselho Nacional de
Desenvolvimento Cient\'ifico e Tecnol\'ogico (CNPq), Brazil under grant
300602/2009-0, by  Funda\c c\~ao para a Ci\^encia e
Tecnologia (FCT), Portugal, under the project No. UID/FIS/04564/2016,
and by NewCompStar, a COST initiative.

\end{document}